\documentclass[aps,pra,superscriptaddress,twocolumn]{revtex4}

\usepackage{graphicx,amssymb,amsmath,color}
\usepackage{nccmath}
\usepackage[pdftex,colorlinks]{hyperref}
\begin{document}
\title{Full Counting Statistics of the momentum occupation numbers of the Tonks-Girardeau gas}
\author{P. Devillard}
\affiliation{Aix Marseille Univ., Universit\'e de Toulon, CNRS, CPT, Marseille, France}
\author{D. Chevallier}
\affiliation{Department of Physics, University of Basel, Klingelbergstrasse 82, CH-4056 Basel, Switzerland}
\author{P. Vignolo}
\affiliation{Universit\'e C\^ote d'Azur, CNRS, Institut de Physique de Nice, France}
\author{M. Albert}
\affiliation{Universit\'e C\^ote d'Azur, CNRS, Institut de Physique de Nice, France}

\begin{abstract}
  We compute the fluctuations of the number of bosons with a given momentum for the Tonks-Girardeau gas at zero temperature. We show that correlations between opposite momentum states, which is an important fingerprint of long range order in  weakly interacting Bose systems, are suppressed and that the full distribution of the number of bosons with non-zero momentum is exponential. The distribution of the quasi-condensate is however quasi Gaussian. Experimental relevance of our findings for recent cold atoms experiments are discussed.
\end{abstract} 

\maketitle

\section{Introduction}
\label{sec_intro}

Ultra-cold atom experiments represent now an established playground to test theories of many-body physics and mimic solid state strongly correlated systems \cite{BlochDalibardZwerger,Lewenstein2012} with an incredible accuracy. One-dimensional systems can be routinely achieved by confining atoms along transverse directions \cite{Caza2011,Guan2013,AltmanDemlerLukin} with the possibility to monitor the interaction strength and the temperature at will. In particular, it is possible to span the entire range of the one-dimensional Bose gas from the weakly interacting to the strongly interacting regime. While pair correlations have a tendency to build albeit without forming a true condensate in the weak coupling limit, strong repulsion tends to make the bosons behave more like fermions. This is the celebrated Tonks-Girardeau gas \cite{Girardeau}. Although physical quantities involving diagonal elements of the density matrix such as spatial density correlations \cite{Caza2011} or the real space emptiness formation probability \cite{Korepinetal} are fermion-like, the off-diagonal part behaves very differently. The most common example is the momentum distribution, namely the average occupation number of a state with a given momentum $p$, $\langle N_p \rangle$, which is the so-called Fermi-Dirac distribution for fermions but is completely different for bosons \cite{Lenard,Leggett2006}.

This momentum distribution is a key observable in the field of ultra cold atoms since it is easily obtained experimentally with time of flight images and contains crucial information on quantum correlations, interaction effects and symmetries of the many-body wave function \cite{Decamp2016}. However, as we know from quantum optics, mesoscopic transport or even the physics of phase transitions, the fluctuations around the average are sometimes the most interesting physical signal. This is why the community is now studying higher moments of the momentum occupation number, like its variance $\langle N_p^2\rangle-\langle N_p\rangle^2$, covariance $\langle N_p N_q \rangle$ \cite{Mathey2009,Rigol2011,Bouchoule2012,FangBouchoule,LovasDoraDemlerZarand,LovasDoraDemlerZarand2} or even the full distribution (full counting statistics) \cite{LovasDoraDemlerZarand}. This can be a great help for unraveling different regimes \cite{Mathey2009,FangBouchoule,Dobrz2019,Carcy2019} or to identify exotic phenomena like the dynamical Casimir effect \cite{Jaskula2012} or Hawking radiation for instance \cite{Unruh1981,Balbinot2008,Recati2009,Fabbri2018,Steinhauer2019}.

\begin{figure} 
  \includegraphics[width=\linewidth]{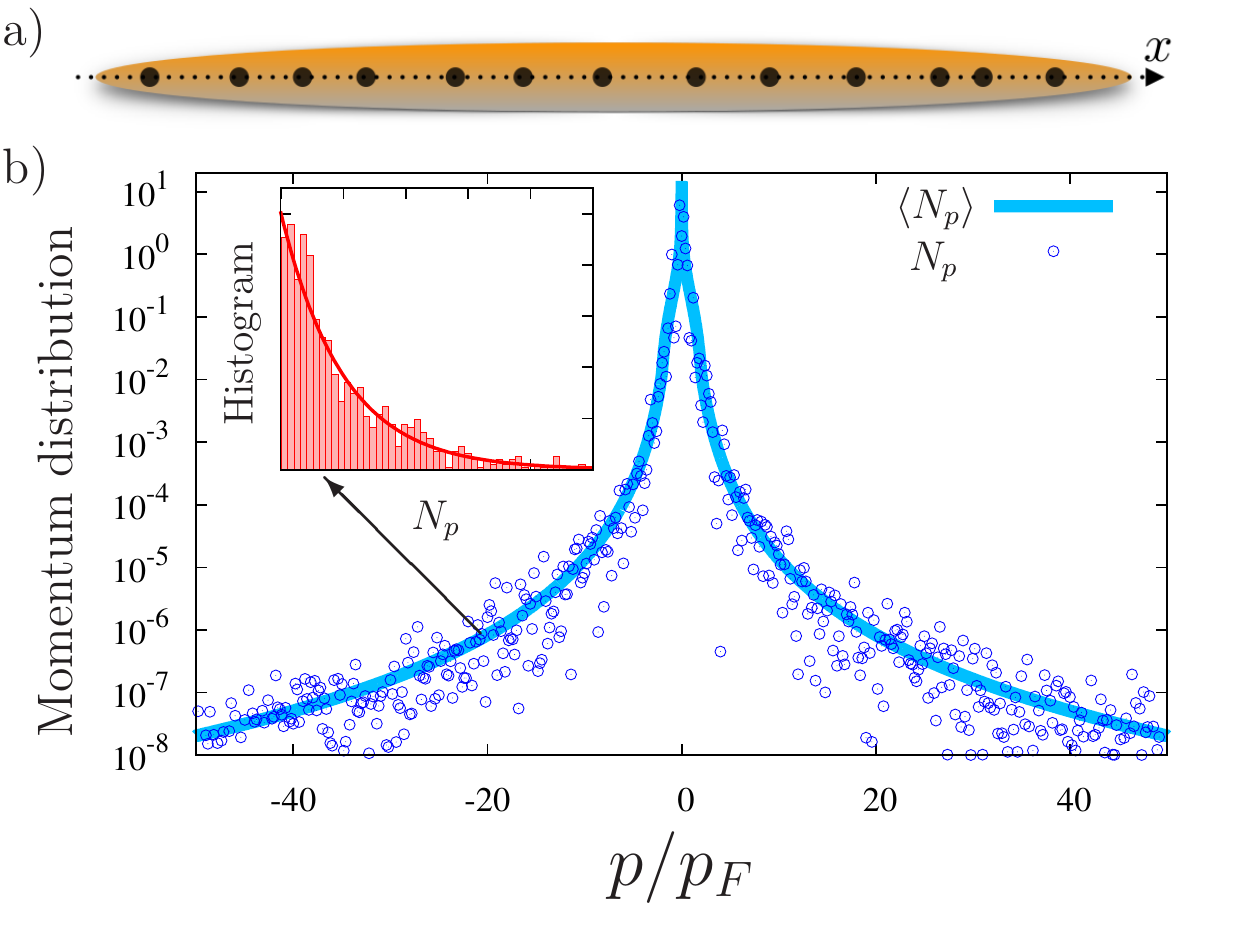}
  \caption{\label{fig_setup} a) One dimensional identical interacting bosons at zero temperature. b) Sketch of the momentum distribution of the 1d Bose gas in the Tonks-Girardeau limit ($p_F=\pi\hbar N/L$). The data represent a single shot measurement whereas the full line is the average $\langle N_p\rangle$. The inset is the full distribution of $N_p$ for a given $p$.}
\end{figure}

In this paper, we study the fluctuations of the momentum occupation number $N_p$ in the Tonks-Girardeau limit at zero temperature for all momenta. This is an extension of the work of Lovas {\it et al.} \cite{LovasDoraDemlerZarand} on the full counting statistics of $N_p$ in the low momentum regime described by bosonization \cite{Giamarchireview,SchoellervanDelft} and the one of Refs. \cite{Mathey2009,Bouchoule2012,FangBouchoule} on the weakly interacting Bose gas. In particular, we show that the full counting statistics of $N_p$ is, for momentum $p$ in almost all regimes, exponential and that the different occupation numbers are uncorrelated. This is in sharp contrast with the weakly interacting regime where Bogoliubov theory predicts positive correlations between opposite momentum states \cite{Mathey2009,Bouchoule2012,BogoliubovShirkov}.  

This article is organized as follows. In Sec. \ref{sec_model} we describe the model and explain the general formalism used to compute the second moment $\langle N_p^2 \rangle$ and the correlations $\langle N_p N_q\rangle$ of the momentum occupation number in terms of Toeplitz matrices. Section \ref{sec_boso} concentrates on intermediate and long wavelength properties. It explains how results from standard bosonization \cite{LovasDoraDemlerZarand} can be retrieved. Section \ref{sec_tan} deals with the opposite limit of large momentum. A small distance expansion of the two-body density matrix enables us to make predictions about the variance of the number of particles with a given momentum. The full probability distribution is also obtained. Section \ref{sec_allp} and \ref{sec_zero} complement our analytical results with numerical calculations of the variance and the correlations for all values of the momentum and for the specific case of the quasi-condensate mode. In the last section, we discuss how our predictions could be tested in realistic experiments and discuss perspectives for future research on other systems along these lines. Technical details can be found in appendices \ref{app_fisher}, \ref{app_tan}, and \ref{app_Bazhanov}.

%------------ Model-------------

\section{Model} 
\label{sec_model}
We consider a gas of $N$ identical bosons living on a strictly one-dimensional segment of length $L$ with periodic boundary conditions. The average density $\rho=N/L$ is constant and we shall mainly be interested in the thermodynamic limit $N \rightarrow \infty$ and $L \rightarrow \infty$ with $N/L$ fixed. However, the formalism also allows to straightforwardly calculate finite size corrections. We focus on the limit of infinite and hardcore repulsion between bosons which is known as the Tonks-Girardeau gas. The Hamiltonian is a limiting case of the Lieb-Liniger model \cite{LiebLiniger} which reads
\begin{equation}
\mathcal H \, = \, -\frac{\hbar^2}{2m}\sum_{i=1}^{N} \frac{\partial^2}{\partial x_i^2}  \, + \, g \sum_{i>j} \delta (x_i - x_j),
\end{equation}
where $x_i$ is the position of the $i^{th}$ bosonic particle of mass $m$, $g$ is the repulsive interaction strength \cite{Olshanii1998}. In the Tonks limit, $g$ is sent to infinity. In this regime, the ground state is constructed by filling all the momentum states up to the Fermi momentum $p_F=\hbar \pi N / L $ while preserving the bosonic statistics as described below. This is the so called regime of fermionization where all physical observables that depend only on density or density correlations are similar to the ones of a perfect gas of fermions \cite{Girardeau}. However, quantum statistics is crucial whenever off-diagonal elements of the density matrix are involved in an observable and this is in particular the case of the momentum distribution \cite{Lenard,Leggett2006}. This quantity, $\langle N_p\rangle$, is the Fourier transform of the one-body density matrix

\begin{eqnarray}
  \label{eq_md}
  \langle N_p\rangle &=&\iint  e^{-i p (x-x')/\hbar}\rho_1(x,x')\,dxdx', \\
  \rho_1(x,x')&=&\int  \Psi^*(X)\,\Psi (X')\, dx_2\cdots dx_N,
\end{eqnarray}
where $X=(x,x_2,...,x_N)$, $X'=(x',x_2,...,x_N)$ and $\Psi(x_1,...,x_N)$ is the many body wave function of the system. It corresponds to the average number of bosons in a state with momentum $p$ and therefore is proportional to the probability of finding a particle with momentum $p$ in an actual experiment. For the Tonks-Girardeau gas considered in this paper, its shape is represented on Fig. \ref{fig_setup} (thick blue line) which has obviously no relation with the one of a perfect Fermi gas (Fermi-Dirac step function at zero temperature). Therefore, many important informations are accessible from this observable such as the quasi-condensate fraction or the symmetry of the wave function. In the Tonks regime, it is known to display several interesting properties. First, the average occupation of the ground state ($p=0$ here) is proportional to $\sqrt{N}$ \cite{Lenard,ForresterFrankelGaroniWitte} and not $N$ like in a weakly interacting Bose gas, signaling the absence of Bose-Einstein condensation in one dimension in the presence of strong interactions. At low momentum, namely for $p\ll p_F$, the momentum distribution decays as $1/\sqrt{p}$ while for $p\gg p_F$ it decays as $p^{-4}$. This latter behavior is universal as long as particles have contact interactions and does not depend on quantum statistics nor on the interaction strength. The coefficient in front of this power law, however, strongly depends on these parameters and is called the Tan contact \cite{Tantheoretical,Tanexperimental}. Some of these features are illustrated on Fig. \ref{fig_std} (thin black line and dashed lines on the main panel).

However, the momentum distribution is only an average quantity. In an experiment, shot to shot fluctuations (blue circles on Fig. \ref{fig_setup}) may be an incredible source of information as it was pointed out by R. Landauer in his famous quote ``the noise is the signal''. With the important advances in the field of single atom detection \cite{Ott2016}, fluctuations around the average will be an additional channel for collecting precious information about the physical properties of quantum liquids but there seems to be very little information about them in the literature. For example, the variance, $\langle N_p^2 \rangle - \langle N_p\rangle^2$ is not known in general. Recently, Lovas {\it et al.} \cite{LovasDoraDemlerZarand} have calculated the probability distribution of the momentum occupation, but only in the long wavelength limit, using bosonization \cite{Caza2011}. They have found that $N_p$ is distributed exponentially in this regime for $p\ne 0$ and that $N_0$ follows a Gumbel distribution for weak interactions. In the opposite limit of large momentum or for strong interaction, nothing is known for the moment. It is the purpose of this paper to answer these questions. We now explain how to compute the variance, the covariance and the full distribution of $N_p$ for a Tonks-Girardeau gas at zero temperature.

In order to calculate the fluctuations of $N_p$, we shall need the two-body density matrix, defined as
\begin{eqnarray}\label{eq_rho2_def}
\rho_2(x,u;y,w) \, &=& \, 
 \int\!\int ...\int 
\Psi^*(x,u,x_3,...,x_N) \nonumber \\
&\,\,& \,\,\, \Psi(y,w,x_3,...,x_N)\, dx_3 \, ... \, dx_N, 
\end{eqnarray} 
with $\Psi$ the ground state wave-function for periodic boundary conditions \cite{Girardeau} 
\begin{equation}
\label{eq_psi}
\Psi(\{x_i\}) \, = {1 \over \sqrt{N! L^N}} 
\prod_{1 \leq j < k \leq N} \!\!
\vert  e^{i {2 \pi \over L} x_j} - e^{i {2 \pi \over L} x_k} \vert.
\end{equation}
Following the steps done in Ref. \cite{ForresterFrankelGaroniWitte} for the one-body density matrix, we apply their method to the two-body density matrix and cast $\rho_2(x,u;y,w)$ as a determinant of a Toeplitz matrix. Technical details are given in Appendix \ref{app_fisher}. This results in 
\begin{equation}
\label{eq_rho2}
\rho_2(x,u;y,w) =\frac{1}{L^2} 
\vert e^{i\theta_u} -  e^{i \theta_x} \vert \,
\vert e^{i \theta_w} -  e^{i \theta_y} \vert \,
 {\rm det}(\Gamma_{i,j}),
\end{equation}
with $\theta_x  = 2 \pi x/L$ ($\theta_y$, $\theta_u$ and $\theta_w$ are defined in a similar fashion) and where ``${\rm det}$'' denotes the determinant of the matrix $\Gamma$ of elements $\Gamma_{i,j}$. This matrix is of Toeplitz type which means that $\Gamma_{i,j}$ is a function of $n = i-j$ only. In addition, $\Gamma$ is also hermitian. Explicitly
\begin{equation}\label{eq_gamma}
\Gamma_n = \, \int_0^{2 \pi} F(\theta) \, e^{i n \theta} {d \theta \over 2 \pi},
\end{equation}
where
\begin{eqnarray}\label{eq_F}
F(\theta) \, &=& \, 16 \, 
\biggl\vert 
\sin \Bigl({\theta - \theta_x \over 2}\Bigr)\,
\sin \Bigl({\theta - \theta_y \over 2}\Bigr)\, \nonumber \\
&\, \, & \,\,\,
\sin \Bigl({\theta - \theta_u \over 2}\Bigr)\,
\sin \Bigl({\theta - \theta_w \over 2}\Bigr)\,
\biggr\vert.
\end{eqnarray} 

Finally, the second moment $\langle N_p^2 \rangle$ can be calculated by taking a double Fourier transform of $\rho_2(x,u;y,w)$. The covariance, namely the correlations between occupation numbers at different momenta $p$ and $q$ is written as 
\begin{equation}
\label{eq_Npq}
\langle N_p N_q \rangle  =  \int_{[0,L]^4}\!\!\!\!\!\!\!\! e^{i\frac{p(y-x)}{\hbar}} e^{i\frac{q (w-u)}{\hbar}}\rho_2(x,u;y,w)\, dx \, dy \, du \, dw. 
\end{equation}

Following this recipe, we will now evalute these quantities analytically at low momentum in Sec. \ref{sec_boso} and large momentum in Sec. \ref{sec_tan} and numerically in Sec. \ref{sec_allp} for any momentum. The reader not so interested in technical details may want to go directly to Sec. \ref{sec_allp}. Finally, the quasi-condensate case ($p=0$) is treated apart in Sec. \ref{sec_zero}.

%-------- Bosonization -----------
\section{Fluctuations in the hydrodynamic regime} 
\label{sec_boso}
In this section, we explain how to retrieve the findings of Ref. \cite{LovasDoraDemlerZarand} on the full distribution of $N_p$ but also compute the covariance in the low momentum regime. Instead of standard bosonization, that is commonly used to describe the physics at low energy, we develop an alternative and more general approach based on asymptotic properties of Toeplitz matrices. At low momentum, $p \ll p_F$ but $p\neq 0$, or large distances compared to $\xi=L/N$, the mean interparticle distance, we show that 

\begin{equation}\label{eq_cov}
  \langle N_p N_q \rangle=(1+\delta_{p,q})\langle N_p\rangle \langle N_q\rangle.
\end{equation}

To do so, we first compute the two-body density matrix in the limit of low momentum as explained in Appendix \ref{app_fisher}. Our calculation, based on the theory of Fisher-Hartwig singularities \cite{Ehrhardt,DeiftItsKrasovsky}, not only reproduce the standard bosonization approach \cite{LovasDoraDemlerZarand} but also allows to compute the numerical prefactor that is in general not possible to obtain. The density matrix reads

\begin{align}\label{eq_rho2_boso}
  \rho_2(x,u;y,w) =& \frac{2N}{L^2} \rho_\infty^2
  \vert e^{i\theta_w}-e^{i\theta_u} \vert^{-\frac{1}{2}}
  \vert e^{i\theta_w}-e^{i\theta_x} \vert^{-\frac{1}{2}}  \nonumber \\
  &\times\vert e^{i\theta_w}-e^{i\theta_y} \vert^{\frac{1}{2}}
  \vert e^{i\theta_y}-e^{i\theta_u}  \vert^{-\frac{1}{2}}  \nonumber \\
  &\times\vert e^{i\theta_y}-e^{i\theta_x} \vert^{-\frac{1}{2}}
  \vert e^{i\theta_u}-e^{i\theta_x} \vert^{\frac{1}{2}},
\end{align}
with $\rho_\infty=G(3/2)^4/\sqrt{2}$ and $G$ is the Barnes function \cite{Grasd}. In addition, if all distances are also much smaller than $L$ we obtain an expression that only depends on terms like $|u-w|$ which is given in Appendix \ref{app_fisher}.

Having determined the two-body density matrix for distances larger than $\xi$, we need to assess the behavior of $\langle N_p^2 \rangle$ and $\langle N_p N_q \rangle$. We start with the former case and notice that due to the oscillatory behavior of the integrand, the integral is dominated by contributions where $p(y-x)/\hbar$ and $p(w-u)/\hbar$ (direct term) or $p(y-u)/\hbar$ and $p(w-x)/\hbar$ (exchange term) are smaller or of order one. We therefore consider configurations in real space where pairs of coordinates are separated by a distance of order $\hbar/p$. By analogy with classical electrodynamics, or to use a more sophisticated language, in the Coulomb gas formulation of the Tonks-Girardeau gas \cite{Fendley1995}, we call these pairs dipoles. Moreover, in the thermodynamic limit, it is very unlikely that two dipoles overlap since their size is typically of order $\hbar/p\ll L$. It is then reasonable to assume that the dipoles are well separated and to simplify the expression of the two-body density matrix to $\rho_2(x,u;y,w) \, \simeq \, \frac{2N}{L^2}  \rho_\infty^2 |e^{i\theta_w}-e^{i\theta_u}|^{-\frac{1}{2}}|e^{i\theta_y}-e^{i\theta_x}|^{-\frac{1}{2}}$ in the direct term and a similar expression for the exchange term. In the approximation of the dilute gas of dipoles, the direct and the exchange terms give the same contribution and Eq. (\ref{eq_Npq}) factorizes to

\begin{align}
\langle N_p^2\rangle=&\,\frac{2 N}{L^2} \rho_\infty^2 \int_0^L \frac{e^{i\frac{p}{\hbar}(y-x)}}{\sqrt{|\sin(\frac{\pi(y-x)}{L})|}} d(y-x)\nonumber \\
&\times\int_0^L \frac{e^{i\frac{p}{\hbar}(w-u)}}{\sqrt{|\sin(\frac{\pi(w-u)}{L})|}} d(w-u).
\end{align}
Here, we recognize twice the square of the momentum distribution (see Eq. (34) of Ref. \cite{ForresterFrankelGaroniWitte} for instance) in the small momentum limit $\langle N_p \rangle=\frac{\sqrt{N}}{L}\rho_\infty\int_0^L  |\sin(\pi x/L)|^{-\frac{1}{2}} dx$. This completes the proof of $\langle N^2_p\rangle=2 \langle N_p\rangle^2$. Note that corrections to this approximation can be calculated by taking into account interactions between dipoles. This can be done by expanding (in Eq. (\ref{eq_rho2_boso2})) $\sqrt{|u-x||w-y|}/\sqrt{|w-x||y-u|}\simeq 1+(w-u)(y-x)/(u-x)^2$ in the direct term (the calculation is similar for the exchange term) but this yields positive corrections of the form $\langle N_p\rangle^2 /p$ which are sub-dominant since $\hbar/L \ll p\ll p_F$ in the thermodynamic limit.

We now compute the covariance $\langle N_p N_q \rangle$ using the same procedure. The direct term gives obviously $\langle N_p\rangle\langle N_{q}\rangle$ whereas the exchange term is a bit more subtle to analyze and reads $\langle N_p N_{q}\rangle_{ex}=\int e^{i\frac{p}{\hbar}(y-u)}e^{i\frac{p}{\hbar}(w-x)}e^{i\frac{(p-q)}{\hbar}(u-w)}\rho_2(x,u;y,w)\, dx \, dy \, du \, dw$. Using the same arguments as before, the two-body density matrix factorizes and no longer depends on $(u-w)$, which due to the presence of the third exponential factor, yields a factor $\delta(p-q)$ in the thermodynamic limit. It is therefore equal to zero for $p\ne q$. Putting pieces together we prove Eq. (\ref{eq_cov}) which suggests that $N_p$ is distributed exponentially. Indeed, pushing forward the dilute gas of dipoles approach, we obtain, for all integers $n$, $\langle N^n_p\rangle=n! \langle N_p \rangle^n$, which is the signature of an exponential distribution 
\begin{equation}\label{eq_exp}
  P(N_p)=\exp(-N_p/\langle N_p\rangle)/\langle N_p \rangle.
\end{equation}
This is precisely the result obtained in \cite{LovasDoraDemlerZarand} using bosonization. However, we will show in the next section that this result is also valid beyond the hydrodynamic regime.

%---------- Tan regime -------------
\section{Short wavelength fluctuations} 
\label{sec_tan}

We now turn to the regime of large momentum, $p \gg p_F$, and extend the previously known results Eqs. (\ref{eq_cov}) and (\ref{eq_exp}). In other words, we demonstrate that $N_p$ is also exponentially distributed with no correlations in the high momentum regime. To prove this, we study the behavior of $\rho_2(x,u;y,w)$ for small $|y-x|$ and $|w-u|$ similar to the short distance expansion of one-body density matrix expansion $\rho_1 (x) \, = \, \rho_1(0) + a x^2 + b \vert x \vert^3 + ...$ \cite{Lenard,Vaidya1979,MinguzziVignoloTosi}. We recall that in the case of the average momentum distribution $\langle N_p\rangle$, the leading term giving the so-called $p^{-4}$ contribution comes from the Fourier transform of $|x|^3$. Indeed, the two first contributions give zero for symmetry reasons and the remaining terms are subdominant in the large $p$ regime. This comes from Watson's lemma \cite{Lighthill} which states that if a function $f(z)$ behaves as $|z-a|^\alpha$ in the vicinity of $a$, then, for large $p$, to leading order in $p$, $\int_{-\infty}^{+\infty} e^{ipz}f(z-a)\,dz=2f(a)e^{ipa}\Gamma(\alpha+1)\cos[\frac{\pi}{2}(\alpha+1)]\,p^{-(\alpha+1)}$. We will see that the situation is similar for the second moment of the distribution. 

Although it is technically possible to perform a cumulant expansion of ${\rm det} (\Gamma_n)$, we shall not pursue this route. We rather use the development by Lenard \cite{Lenard}. This formal series is an expansion of the two-body bosonic density matrix in terms of the fermionic ones and reads
\begin{eqnarray}
&\!\!& {\rho_2} (x,u;y,w) =  
 {\rm sgn}(u-x) \, {\rm sgn}(w-y) \,
\Biggl\lbrack
\langle x,u \vert \rho_F \vert y,w \rangle \nonumber \\
&\,& +{(-2) \over 1!} 
\int_J \langle x,u,x_3 \vert \rho_F \vert y,w,x_3 \rangle \, 
dx_3 + \cdots  \nonumber \\ 
& \, &  + {(-2)^n \over n!} 
\int_J \! \int_J \cdots \int_J dx_3 ... dx_{n+2}  \nonumber \\
&\! &
\times \langle x,u,x_3,...,x_{n+2} \vert \rho_F \vert y,w,x_3, ... ,x_{n+2} \rangle
 +\cdots\Biggr\rbrack  ,
\end{eqnarray}
where the interval $J$ is defined as $J \, \equiv \, \lbrack x \, , \, y \rbrack \, \cup \lbrack u \, , \, w \rbrack$ and $\rho_F$ is the fermionic density matrix. The $m$-body fermionic density matrix reads
\begin{align}
 & \langle  x,u,x_3,...,x_{m} \vert \rho_F \vert y,w,x_3, ... ,x_{m} \rangle = L^{-m}\times 
  \nonumber \\
 & \!\!\!\!\!\!\!\!
\left\vert
\begin{matrix}
f(y-x) & f(w-x) & f(x_3-x) & ... & f(x_{m}-x) \cr
f(y-u) & f(w-u) & f(x_3-u) & ... & f(x_{m}-u) \cr
f(y-x_3) & f(w-x_3) & f(x_3-x_3)  & ... & f(x_{m}-x_3) \cr
... & ... & ... & ... &  \cr
f(y-x_{m}) & f(w-x_{m}) & f(x_3-x_{m}) & ... & N
\end{matrix}
\right\vert \nonumber, \\
\end{align}
with $f(z) \, \equiv \, {\sin (N\pi z/L) \over \sin( \pi z/L)}$, for  $z \not= 0$ and $f(0) = N$. Although it is possible to compute all terms for finite $N$, we directly take the thermodynamic limit for the sake of simplicity. Moreover, it is again sufficient to consider dilute dipole configurations since clusters of more than two points give subdominant contributions. This time, it is simply related to the fact that the density matrix vanishes as a power law when two spatial coordinates approach each other (in the Coulomb gas formulation of the Tonks-Girardeau gas, these configurations are strongly penalized by Coulomb repulsion). This can be easily understood by looking at the functional dependence of the many-body wave function Eq. (\ref{eq_psi}). In this limit, we have computed this expansion explicitly up to seventh order in $|u-x|$ and $|w-y|$ as it was necessary to obtain the relevant contribution. All the terms are collected in Appendix \ref{app_tan}.

In order to compute the variance and the correlations, we use a similar dipole decomposition of the Fourier transform with a direct term corresponding to $|x-y|\ll \xi$ and $|w-u|\ll\xi$ with $|u-x|\gg\xi$ and an exchange term with $|x-w|\ll \xi$ and $|y-u|\ll\xi$, also with $|u-x|\gg\xi$. It turns out that, as long as $|u-x|\gg\xi$, the expansion is independent of $(u-x)$, which makes the calculation of the Fourier transform rather easy. The expansion for the direct term is of the form 
\begin{equation}\label{rho2_exp}
\rho_2(x,u;y,w) =  \sum_{n=0}^{\infty} \sum_{m=0}^n A_{m,n} \left| \frac{y-x}{\xi} \right|^{m} \left\vert \frac{w-u}{\xi} \right\vert^{n-m}.
\end{equation}
When looking carefully at the different terms, it turns out that the relevant term is $A_{3,6} \vert (y-x)/\xi \vert^3 \, \vert (w-u)/\xi \vert^3$. Performing the same expansion for the exchange term and lumping the two expansions together yield immediately $\langle N_p^2 \rangle \, = \, 2 \langle N_p \rangle^2$, with $\langle N_p \rangle =  C p^{-4}$ and $C = {4 \over 3\pi^2}p_F^4$ since the direct and the exchange contributions are identical. However, for the correlations, the exchange contribution vanishes for the same reason as in Sec. \ref{sec_boso}. Therefore, we also find that no correlation exists between different momenta in this limit. In particular, $N_p$ and $N_{-p}$ are not correlated as opposed to what happens in the weakly interacting regime \cite{BogoliubovShirkov}.

The above analytical part of the calculation can be generalized to the $n$-body density matrix $\rho_n(z_1,z_2, ... , z_n ; s_1, s_2, ... , s_n)$. This gives access to the $n^{{\rm th}}$ moment of $N_p$, $\langle N_p^n \rangle$, resulting in $\langle N_p^n \rangle \, = \, n! \Bigl({C \over p^4}\Bigr)^n$, in the limit $p \gg p_F$. The knowledge of all the integer moments $\langle N_p^n \rangle$ enables \cite{Stieltjesmomentproblem,BarrySimon} us to reconstruct the probability distribution $P(N_p)$ which is therefore exponential. 

\begin{figure} 
  \includegraphics[width=\linewidth]{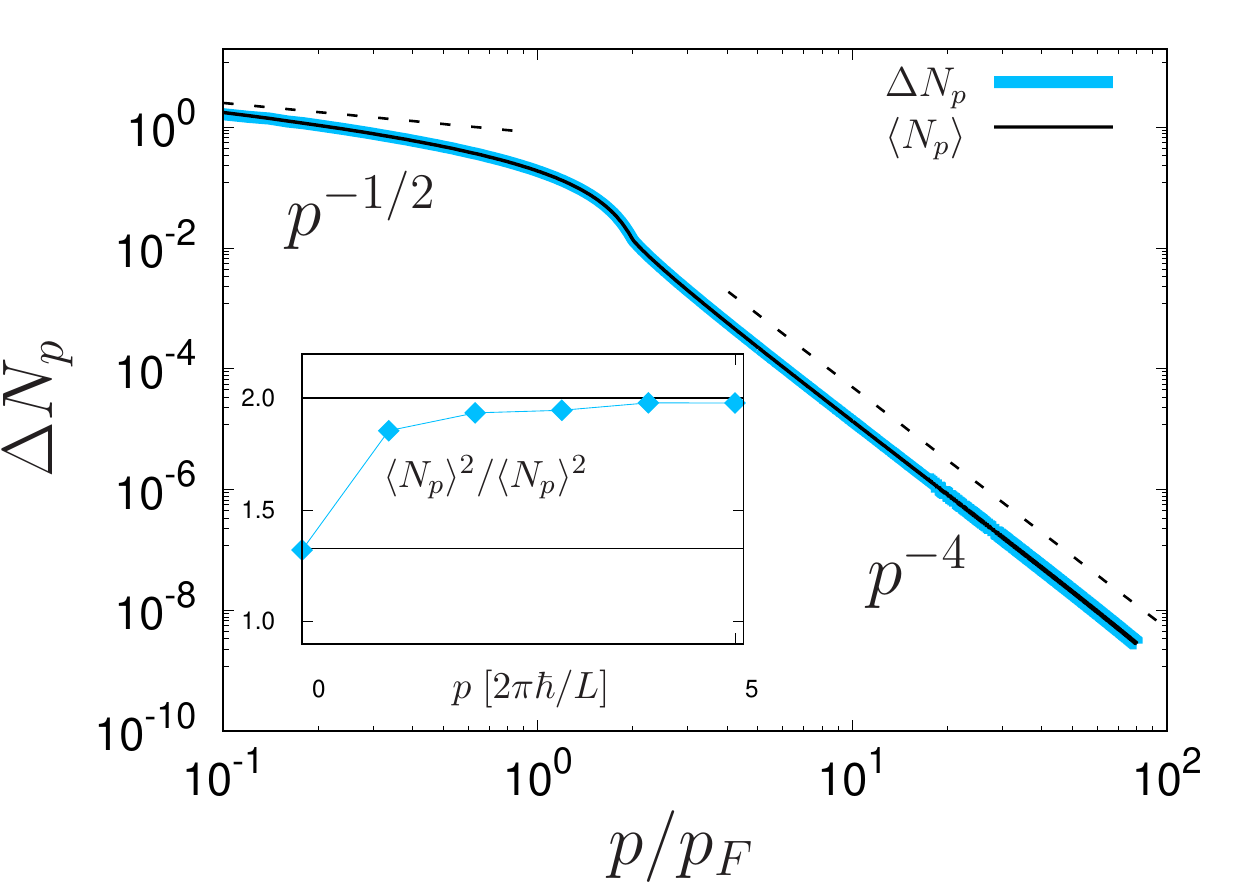}
  \caption{\label{fig_std}Standard deviation $\Delta N_p=\sqrt{\langle N_p^2\rangle-\langle N_p\rangle^2}$ (blue thick solid line) and average number of bosons  $\langle N_p \rangle$ (black thin solid line) as a function of $p$ for $N=100$ bosons (blue thick solid line). Dashed lines are the limiting cases (see text). The inset shows the same data in linear scale in a slightly different form. The ratio $\langle N^2_p \rangle/\langle N_p \rangle^2$ is plotted as a function of the momentum $p$ in units of $2\pi\hbar/L$. The horizontal black lines correspond to the two limiting values $1.33$ and $2$ (see Eqs. (\ref{eq_N02}) and (\ref{eq_cov})). The continuous line is merely a guide to the eyes.}
\end{figure}

%---------- Intermediate regime ----------

\section{Intermediate regime} 
\label{sec_allp}
\begin{figure} 
  \includegraphics[width=\linewidth]{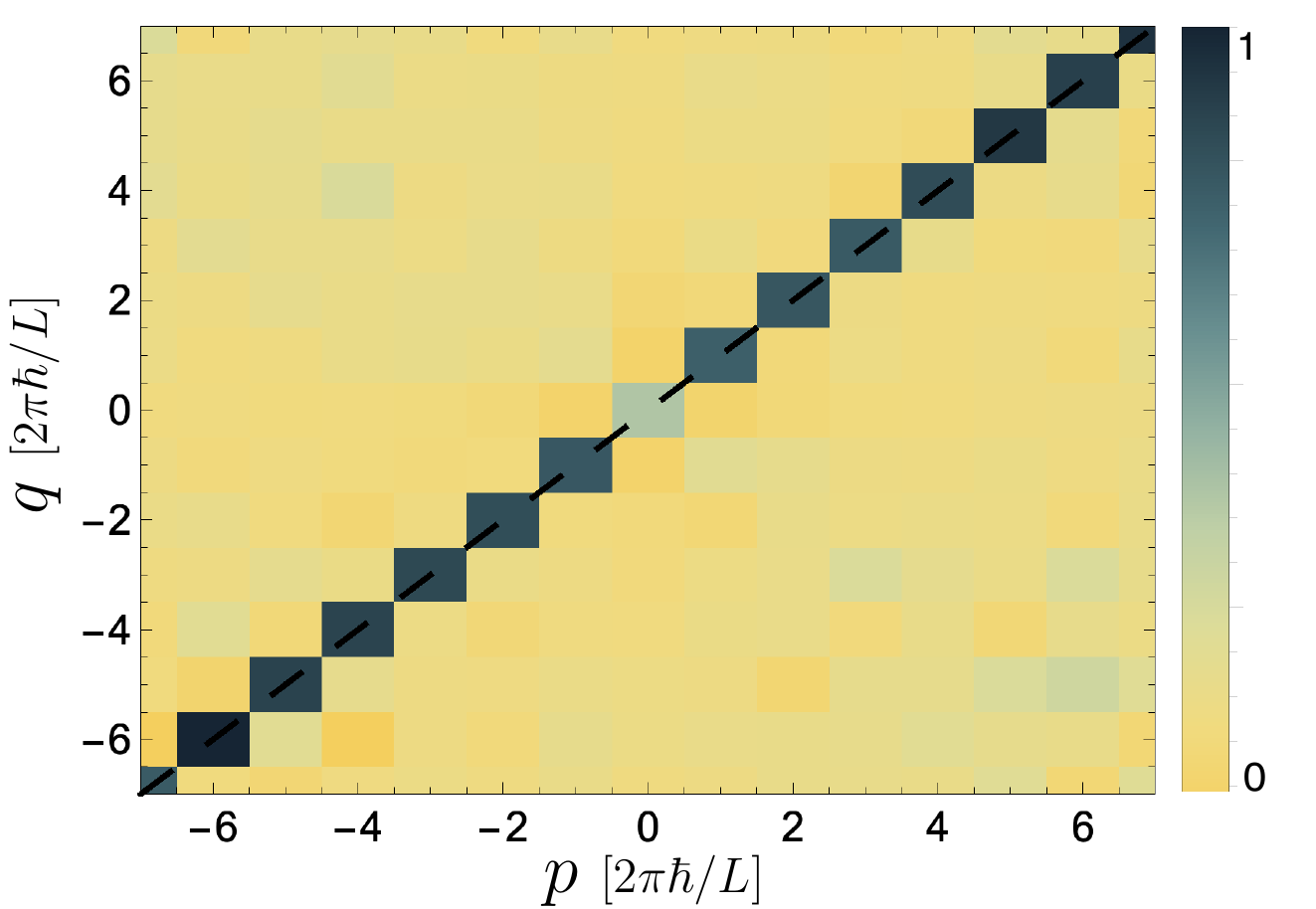}
  \caption{\label{fig_corr}(Color online) Normalized correlations between different momentum occupation numbers $\langle N_p N_{q}\rangle/\langle N_p\rangle\langle N_q\rangle-1$ for $N=100$ bosons as a function of $p$ and $q$. The dashed line indicates the diagonal $p=q$. A cut along the diagonal is visible in the inset of Fig. \ref{fig_std}.}
\end{figure}

We have now proven that the occupation number of a state with momentum $p$ is exponentially distributed according to Eq. (\ref{eq_exp}) and that occupation numbers with different momentum are uncorrelated for small but non zero ($\hbar/L\ll p\ll p_F$) and large momenta ($p\gg p_F$). It is then natural to wonder if this statement is correct for intermediate momentum. In that case, we have computed numerically the variance and covariance of $N_p$ using Eqs. (\ref{eq_rho2}) and (\ref{eq_Npq}). Our results are presented on Fig. \ref{fig_std} and Fig. \ref{fig_corr} for the standard deviation $\Delta N_p=\sqrt{\langle N_p^2\rangle-\langle N_p \rangle^2}$ and the normalized correlations $\langle N_p N_q\rangle/\langle N_p\rangle\langle N_p\rangle-1$ respectively. 

As can be seen on Fig. \ref{fig_std}, the numerically obtained curves $\Delta N_p$ and $\langle N_p \rangle$ are almost indistiguishable from each other for all values of $p/p_F$, not only in the large and small momentum regimes. The inset in Fig. \ref{fig_std} shows deviations to this law that will be discussed in the next section. Although it is not a proof, it is a strong evidence that the equation $\langle N_p^2 \rangle \, = \, 2 \langle N_p\rangle^2$ is valid for any momentum $p$, as long as $p$ is not too close to zero, as discussed in Sec. \ref{sec_zero} below. It is therefore reasonable to believe that $N_p$ is distributed exponentially for any value of $p\neq 0$. This has the important consequence that for a Tonks-Girardeau gas, the relative fluctuations of $N_p$ never vanish in the thermodynamic limit. They are always equal to the signal itself. This is schematized in Fig. \ref{fig_setup}.

Concerning the correlations, one can also observe on Fig. \ref{fig_corr} that they exist only for $p=q$ in agreement with Eq. (\ref{eq_cov}). Indeed, only one straight line on the color map $\langle N_p N_q\rangle/\langle N_p \rangle \langle N_q\rangle-1$ as a function of $p$ and $q$ is visible, the rest of the color map being zero. This is in sharp contrast with the physics of a weakly interacting Bose gas discussed in Refs. \cite{Mathey2009,Bouchoule2012,FangBouchoule}, where, for instance, correlations between $p$ and $-p$ are clearly visible. This is not really a surprise since these pair correlations stem from the existence of a condensate and are the hallmark of long range coherence. They basically emerge from the low energy excitations of this system that are phonons which are quasi-particles with equal weight of opposite momentum components whereas in the Tonks-Girardeau gas, the low energy excitations are particle-hole like and independent from each other.  

%-------- Quasi Condensate ----------

\section{Quasi-condensate mode}
\label{sec_zero}
So far, we have focused on the statistical distribution and correlations of momentum states occupation numbers with non-zero momentum. As discussed in \cite{LovasDoraDemlerZarand}, the quasi-condensate mode which has zero momentum must be treated differently. Using arguments based on Bogoliubov theory in the weak coupling regime, Lovas {\it et al.} have explained that the distribution of $N_0$ was of Gumbel type \cite{Gumbel}. However, at larger coupling (when the Luttinger parameter $K$ approaches one in Fig. 3 of \cite{LovasDoraDemlerZarand}), important deviations from this prediction are visible. In the following, we briefly discuss how this problem is related to other models that have been studied in the literature and discuss some important results such as the variance and the shape of the distribution of $N_0$.

The $n^{th}$ moment of the number of bosons in the zero momentum state $\langle N_0^n \rangle$ is given by the following formula
\begin{widetext}
\begin{equation}
\label{eq_N0n}
\frac{\langle N_0^n \rangle}{(\rho_\infty\sqrt{2N})^n} = \int_0^{2 \pi}\!\!\!\!\cdots\int_0^{2 \pi}
  \!\!\!\!\prod_{1 \leq i  < j \leq n} \vert 4 \sin \Bigl({\theta_i - \theta_j \over 2}\Bigr) \sin \Bigl({\theta^{\prime}_i - \theta^{\prime}_j \over 2}\Bigr)\vert^{\alpha}  \Biggl(\prod_{i=1}^n \prod_{j=1}^n \vert 2 \sin\Bigl({\theta_i - \theta^{\prime}_j \over 2}\Bigr)\vert^{\alpha}\Biggr)^{-1} {d \theta_1 \over 2 \pi}   \cdots  {d \theta_n \over 2 \pi} \,\, {d \theta^{\prime}_1 \over 2 \pi}   \cdots {d \theta^{\prime}_n \over 2 \pi},
\end{equation}
\end{widetext}
with $\alpha = 1/2$. This kind of expression shows up in other physical problems and has been studied in different contexts. For instance, if divided by $n!^2$, it can be interpreted as the canonical partition function of a neutral two-component Coulomb gas with $2n$ (in total) logarithmically interacting charges \cite{Samaj2013}. Then, $\theta_i$ and $\theta^{\prime}_j$ are the positions of the $+$ and $-$ charges on the unit circle respectively. The inverse temperature of the Coulomb gas is $\beta = \alpha$. It is also related to the partition function which describes tunneling through a barrier of an interacting spinless Luttinger liquid with attractive interactions \cite{KaneFisher} and interaction parameter $g=4$ in the notations of Ref. \cite{KaneFisher}. The gas is in the disordered phase, at a temperature $T$ well  above the Kosterlitz-Thouless transition $T_{KT}$, which occurs at $\beta = 1/T_{KT} = 2$ in their units. Finally, this problem of finding the full distribution of $N_0$ is closely related to the full counting statistics of the average interference patterns between two Bose condensates \cite{Gritsevetal1,Gritsevetal2}. However, in the case of Refs. \cite{Gritsevetal1,Gritsevetal2}, there are two condensates, each one having a Luttinger parameter $K$. Consequently, Eq. (\ref{eq_N0n}), giving $\langle N_0^n \rangle$ translates to the same problem they studied but with $\alpha = {1 \over 2 K}$ and not $1/K$. For the Tonks-Girardeau gas, $\alpha = 1/2$ and we can thus use the results derived in Ref. \cite{Gritsevetal1}, with $K=2$, instead of $K=1$, as one might naively think. Therefore, most of the results about the distribution of $N_0$ are available in the references mentioned above. In particular, using previous work by Bazhanov {\it et al.} \cite{Bazhanovetal}, the authors of Ref. \cite{Gritsevetal1} were able to obtain a distribution related to $P(N_0)$ exactly. 

We now discuss several simple results, namely the two first moments of the distribution and its shape. The average value of $N_0$ was calculated in \cite{ForresterFrankelGaroniWitte} and reads

\begin{equation}
  \langle N_0\rangle=\frac{\sqrt{2\pi}}{[\Gamma(3/4)]^2}\rho_\infty\sqrt{N},
\end{equation}
while the second moment can be evaluated numerically from Eq. (\ref{eq_N0n}) and gives \cite{noteint}
\begin{equation}
  \langle N^2_0\rangle\simeq 1.33\,\langle N_0\rangle^2,
  \label{eq_N02}
\end{equation}
which shows that $N_0$ is no longer exponentially distributed: $\langle N^2_0\rangle\neq 2\langle N_0\rangle^2$. Nevertheless the fluctuations of $N_0$ are proportional to its average and therefore do not disappear either in the thermodynamic limit as it has also been noticed in lattice systems \cite{Rigol2011}. The prefactor in Eq. (\ref{eq_N02}) is smaller than two which means that fluctuations are smaller in the quasi-condensate than in other modes. We associate this to a reminiscent effect of coherence that would reduce fluctuations in the condensate. This result is depicted in the inset of Fig. \ref{fig_std}. At $p=0$ it can be seen that the prediction of Eq. (\ref{eq_N02}) is verified (see the lower black horizontal line) and that for $p\neq 0$ the statistics quickly converges to the exponential one.

\begin{figure} 
  \includegraphics[width=\linewidth]{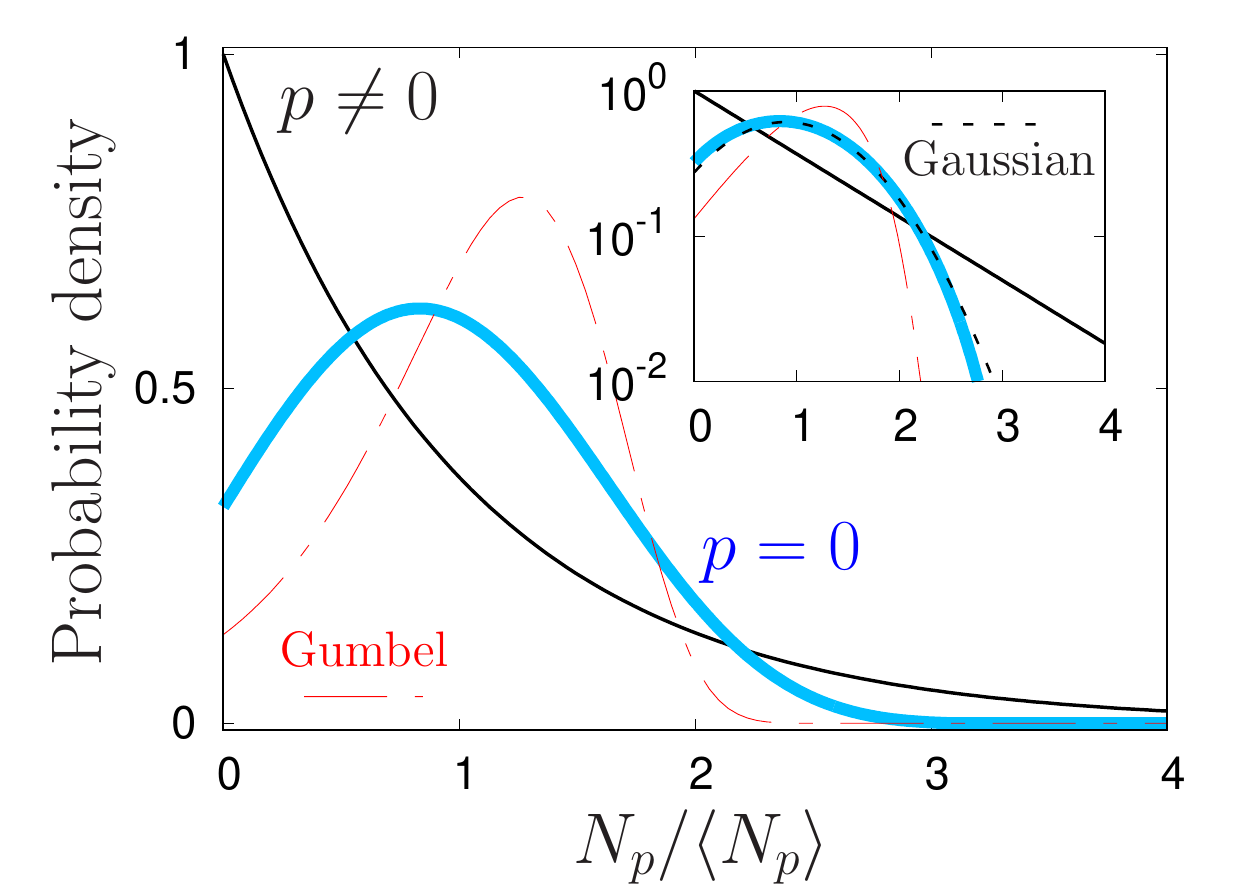}
  \caption{\label{fig_pno} Probability densities of $N_0/\langle N_0\rangle$ (thick blue line) and $N_p/\langle N_p\rangle$ (thin black line) for $p\ne 0$ (but $p\gg\hbar/L$). The inset shows the same data in semi-log scale. The red dotted-dashed curve is the Gumbel distribution \cite{Gumbel} and the black dashed curve is a Gaussian fit for guidance.}
\end{figure}

In addition to the average and the variance, we have access to the full distribution. Following the method employed in Ref. \cite{Gritsevetal1} (see Appendix \ref{app_Bazhanov}) for $K=2$ in their notation, we have calculated the distribution of $N_0$. The result is shown on Fig. \ref{fig_pno} and demonstrates that in the Tonks-Girardeau regime, it is neither exponential nor Gumbel but still contains large fluctuations. Some insight on the full distribution of $N_0$ can also be obtained by looking at the asymptotic behavior of the moments. Using the results of Ref. \cite{Fendley1995}, we obtain for $n\gg 1$

\begin{equation}
  \langle N_0^n \rangle \simeq (\rho_\infty\sqrt{2N})^n \exp \bigl\lbrack {1 \over 2} n \ln n + {\cal O}(n) \bigr\rbrack,
\end{equation}
which can easily be checked to be the asymptotic expression of the moments of a positive Gaussian distributed random variable. This is indeed what is apparent in the inset of Fig. \ref{fig_pno} where we show the probability density in logarithmic scale. This result can also be retrieved analytically by looking at the behavior of spectral determinants, along the lines of Refs. \cite{Gritsevetal1,Gritsevetal2}; see Appendix \ref{app_Bazhanov}. The advantage of this method is that it also permits to obtain information on the behavior of $P(N_p)$ for very small but non-zero momenta $p$ but we leave this for future investigations.

%-----------Conclusion----------

\section{Conclusion and perspectives}
\label{sec_ccl}
In this work, we have proposed a scheme to compute the quantum fluctuations, at zero temperature, of the number of particles $N_p$ with momentum $p$, for the Tonks-Girardeau gas. We have shown analytically in the low ($\hbar/L \ll p \ll p_F$) and high momentum limits ($p\gg p_F$) and have given strong numerical evidences for intermediate values of momentum that $N_p$ is distributed according to an exponential law. In particular, we have demonstrated that the standard deviation of the momentum distribution was equal to its mean value. In addition, we have computed the covariance $\langle N_p N_q\rangle$ and shown that correlations were only visible on the axis $p=q$ and that correlations between $N_p$ and $N_{-p}$ were suppressed contrary to the case of a weakly interacting Bose gas described by Bogoliubov quasi-particles. Finally, the distribution of the quasi-condensate mode at $p=0$ was shown to behave differently as already observed for weak and moderate interaction in \cite{LovasDoraDemlerZarand}. In the Tonks regime, we argued that the tails of its distribution is neither exponential nor Gumbel but rather of Gaussian type. The case of correlations for very small but non-zero momentum ($p\simeq \hbar/L$) is more difficult and is left for future investigations.

Our findings can be relevant for ultra-cold atom experiments where high-order correlation functions in momentum space can be measured, for instance, with time of flight techniques \cite{ClementAspect}. In actual experiments, atoms are generally released from a harmonic trap and the effect of the well potential on the momentum distribution has to be taken into account \cite{ClementAspect,TanLiebLiniger}. Inclusion of finite temperature would also be a natural generalization of this work \cite{Minguzzi2013,Deuaretal,Rigol2015} as well as finite interaction corrections in the regime of large momentum \cite{Lang2017}. Investigating the weak coupling or intermediate coupling of the boson interaction, i.e. using the Lieb-Liniger model \cite{LiebLiniger} would also provide more insight \cite{GangardtShlyapnikov1,GangardtShlyapnikov2,OlshaniiDunjko} on how the quasi-condensate correlations build up \cite{Nandanietal}. Finally, another important lead to follow would be the study of the fermionic counterpart where generalization of random matrix theories \cite{Sutherland1971,Schehr2019}, including off-diagonal contributions of the density matrix  would have to be considered. 

%-----Acknowledgments---------

\section*{Acknowledgments} We would like to acknowledge helpful discussions with D. Cl\'ement, J. Decamp and M. Rigol. The work of D. C. was supported by the Swiss NSF and NCCR QSIT.
 
\appendix 

%-------Appendix I --------
\section{Asymptotic behavior of the two-particle density matrix from determinants with Fisher-Hartwig singularities}
\label{app_fisher}

In this appendix, we derive the expression of the two-body density matrix of the Tonks-Girardeau gas in terms of Toeplitz matrices and compute its long distance approximation using asymptotic properties of these matrices \cite{Ehrhardt,DeiftItsKrasovsky}. 

Starting from Eq. (\ref{eq_rho2_def}) of the main text, inserting the ground state wave-function Eq. (\ref{eq_psi}), and defining $\theta_{x_i} = 2 \pi x_i/L$, we obtain $\rho_2(x,u;y,w)$
\begin{align}
\rho_2  &=  {1 \over N! L^N} \,
\int_0^{2 \pi} \cdots \int_0^{2 \pi} 
\vert e^{i \theta_x} - e^{i \theta_u} \vert
\vert e^{i \theta_w} - e^{i \theta_y} \vert  \nonumber
 \\
&  \times\biggl( \prod_{l=3}^N
\vert e^{i \theta_x} - e^{i \theta_l} \vert
\vert e^{i \theta_y} - e^{i \theta_l} \vert
\vert e^{i \theta_u} - e^{i \theta_l} \vert
\vert e^{i \theta_w} - e^{i \theta_l} \vert
\biggr)& \nonumber \\
& \times\prod_{3 \leq m < n \leq N} \vert e^{i \theta_m} - e^{i \theta_n} \vert^2
dx_3 ... dx_N.
\end{align}
Then using the formulation in terms of a determinant of a Toeplitz matrix, see Ref. \cite{Lenard,Grenander1958}, we use the lemma
\begin{equation}
\begin{array}{l}
\displaystyle{1 \over N!} \int_0^{2 \pi}\!\!\!\!\! \cdots \int_0^{2 \pi} 
\prod_{l=1}^N f(\theta_l)  \, 
\!\!\!\!\!\!\!\!\prod_{1 \leq n < m \leq N}\!\!\!\! \vert e^{i \theta_m} - e^{i \theta_n} \vert^2 \, 
{d\theta_1 \over 2 \pi} \dots {d \theta_N \over 2 \pi} \\
\quad = \, \displaystyle{\rm det}(M), 
\end{array}
\end{equation}
where $M$ is the square matrix with elements $M_{m,n}  = \int_0^{2 \pi} e^{i \theta  (m-n)} f(\theta)   {d \theta \over 2 \pi}$. This lemma follows directly from expressing $\prod_{3 \leq m < n \leq N} \vert e^{i \theta_m} - e^{i \theta_n} \vert^2$ as the square of a Vandermonde determinant, namely 
\begin{equation}
  \prod_{1 \leq n < m \leq N}\!\!\!\!\!\!\! \vert e^{i \theta_m} - e^{i \theta_n} \vert^2  = \sum_{{\cal P},{\cal Q}} \epsilon({\cal P})\epsilon({\cal Q}) \prod_{l=1}^N e^{i \theta_l \lbrack {\cal P}(l) - {\cal Q}(l) \rbrack},
\end{equation}
where ${\cal P}$ and ${\cal Q}$ are permutations of the $N$ integers from $1$ to $N$. $\epsilon({\cal P})$ is the signature of the permutation ${\cal P}$. The sum on ${\cal P}$ runs over all the $N!$ permutations, so as the one on ${\cal Q}$. In our case, we take out the term ${1 \over L^N} \, \vert e^{i \theta_x} - e^{i \theta_u} \vert \vert e^{i \theta_w} - e^{i \theta_y} \vert$ and apply the lemma with $N-2$ instead of $N$ and 
\begin{equation}
  f(\theta) = \vert e^{i \theta_x} - e^{i \theta} \vert\vert e^{i \theta_y} - e^{i \theta} \vert  \vert e^{i \theta_u} - e^{i \theta} \vert  \vert e^{i \theta_w} - e^{i \theta} \vert.
\end{equation}
Since $\vert e^{i \theta_x} - e^{i \theta} \vert = 2 \bigl\vert \sin \bigl({\theta- \theta_x \over 2} \bigr) \bigr\vert$, we obtain Eqs. (\ref{eq_rho2}), (\ref{eq_gamma}), and (\ref{eq_F}) of the main text. 

We now evaluate the large $N$ behavior of the two-body density matrix. In the spirit of Refs. \cite{Lenardpacific,Basor}, we adapt the method used there for the one-body density matrix to the large distance behavior of the two-body density matrix which is governed by the Fisher-Hartwig singularities of the matrix $\Gamma_{i,j}$, in Eqs. (\ref{eq_rho2}) and (\ref{eq_gamma}). We suppose that $x$, $y$, $u$, and $w$ are all separated by a distance larger than $L/N$. Starting from Eq. (\ref{eq_rho2}), we need to evaluate the asymptotic behavior of ${\rm det}(\Gamma_{i,j})$ for large $N$, with $\theta_x$, $\theta_y$, $\theta_u$ and $\theta_w$ larger than $N^{-1}$. The symbol $F(\theta)$ of the Toeplitz matrix $\Gamma_{i,j}$ is given by Eq. (\ref{eq_F}) of the main text and satisfies $\int_0^{2 \pi} \ln \, F(\theta) \, d\theta \, = 0$. This implies that the determinant does not increase nor decays exponentially for large $N$. There are however four distinct Fisher-Hartwig singularities located at $\theta = \theta_x$, $\theta_y$, $\theta_u$ and $\theta_w$. These singularities are all of the same type, a discontinuity of the slope in $F(\theta$); in other words there are four $\alpha$-type singularities in the notations of Ref. \cite{Basor}, with $\alpha = 1/2$. Applying theorems (2) and (3) from Ref. \cite{Basor}, we obtain,
\begin{align}
&{\rm det}(\Gamma_{i,j}) \simeq N G(3/2)^8 \nonumber \\
&\times \vert e^{i \theta_y} - e^{i \theta_x} \vert^{-\frac{1}{2}}
 \vert e^{i \theta_y} - e^{i \theta_u} \vert^{-\frac{1}{2}} 
 \vert e^{i \theta_w} - e^{i \theta_x} \vert^{-\frac{1}{2}}\nonumber\\
& \times\vert e^{i \theta_w} - e^{i \theta_u} \vert^{-\frac{1}{2}}
 \vert e^{i \theta_w} - e^{i \theta_y} \vert^{-\frac{1}{2}}
 \vert e^{i \theta_u} - e^{i \theta_x} \vert^{-\frac{1}{2}},
\end{align}
with $G$ the Barnes function \cite{Grasd}. Now, taking into account the prefactor in Eq. (\ref{eq_rho2}),  
\begin{align}
 & \rho_2(x,u;y,w) \simeq  (N/L^2)\,  G(3/2)^8 \nonumber \\
 & \times\vert e^{i \theta_y} - e^{i \theta_x} \vert^{-\frac{1}{2}}
  \vert e^{i \theta_y} - e^{i \theta_u} \vert^{-\frac{1}{2}}
  \vert e^{i \theta_w} - e^{i \theta_x} \vert^{-\frac{1}{2}}  \nonumber \\
 & \times\vert e^{i \theta_w} - e^{i \theta_u} \vert^{-\frac{1}{2}}
  \vert e^{i \theta_w} - e^{i \theta_y} \vert^{+\frac{1}{2}}
  \vert e^{i \theta_u} - e^{i \theta_x} \vert^{+\frac{1}{2}},
\end{align}
which is Eq. (\ref{eq_rho2_boso2}) in the main text. In order to retrieve the familiar result of bosonization on the infinite line, we suppose that all arguments $x,y,u,w$ are small with respect to $L$, but can be large with respect to $L/N$. This allows to approximate $\vert e^{i \theta_y} - e^{i \theta_x} \vert^{-\frac{1}{2}} \simeq \sqrt{L/2 \pi} \vert y-x \vert^{-\frac{1}{2}}$ and yields
\begin{eqnarray}\label{eq_rho2_boso2}
  &\rho_2(x,u;y,w) = \, N[ G(3/2)^8\,/ (2 \pi L)]
  \vert w-u \vert^{-\frac{1}{2}}
  \vert w-x \vert^{-\frac{1}{2}} \nonumber \\
  &\times \vert w-y \vert^{\frac{1}{2}}
  \vert y-u \vert^{-\frac{1}{2}}
  \vert y-x \vert^{-\frac{1}{2}}
  \vert u-x \vert^{\frac{1}{2}}.
\end{eqnarray}

%---Appendix II -------
\section{Thermodynamic limit of $\rho_2(x,u;y,w)$ for $|u-x| \gg \xi$, at short distances, $|y-x|$ and $|w-u|$ $\ll \xi$}
\label{app_tan}
We give here explicit expressions of the Lenard expansion in the thermodynamic limit,
 in the regime of the dilute gas of dipoles, up to seventh order. These results are simply obtained by computing the determinants in the large $N$ limit. Here we consider the configuration where $x$ and $y$ and $u$ and $w$ constitute the two dipoles ($|y-x|\ll\xi$ and ($|w-u|\ll\xi$)) that are far apart ($|u-x|\gg \xi=L/N$) but it is straightforward to obtain all possible permutations since the bosonic density is symmetric with respect to permutations. In that case, an important simplification comes from the fact that ${\sin (\pi N (u-x)/L)  \over \sin( \pi (u-x)/L)}$ is always of order $1$ and never of order $N$, giving lower powers of $N$. A tedious calculation to the seventh order yields
\begin{equation}
  \rho_2(x,u;y,w) = \frac{N^2}{L^2} {\rm sgn}(u-x) \, {\rm sgn}(w-y) \sum_{n=0}^7 T_n,
\end{equation}
with
\begin{fleqn}
\begin{equation}
T_0 = 1, \quad
T_1 = 0,
\end{equation}
\begin{equation}
T_2 = - {\pi^2 \over 6} (Y^2 + W^2), \quad
T_3 = - {\pi^2 \over 9} (\vert Y\vert^3 + \vert W\vert^3),
\end{equation}
\begin{equation}
T_4 = \Bigl({\pi^4\over 120} + {\pi^2 \over 9}\Bigr) (Y^4 + W^4) \, + \, {\pi^4 \over 36} Y^2 W^2,
\end{equation}
\begin{eqnarray}
T_5 &=&  - {11 \over 1350} \pi^4 (\vert Y \vert^5 + \vert Y \vert^5)\nonumber \\
& &- {\pi^4 \over 54} (\vert Y \vert^3 W^2 + Y^2 \vert W \vert^3),
\end{eqnarray}
\begin{eqnarray}
  T_6 &=&  \Bigl({\pi^2 \over 9}\Bigr)^2 \vert Y \vert^3 \, \vert W \vert^3 - \Bigl({\pi^6 \over 5040}+ {11 \over 450} \pi^4\Bigr) (Y^6 + W^6) \, \nonumber \\
& & - \Bigl( {\pi^6 \over 720} + {\pi^4 \over 54}) (Y^4 W^2 + Y^2 W^4),
\end{eqnarray}
\begin{eqnarray}
  T_7 &=& {61 \over 264600} \pi^6 \, (\vert Y \vert^7 + \vert W \vert^7)\nonumber\\
  & & +  {11 \over 1800} \pi^6 \, (\vert Y \vert^5 W^2 + Y^2 \vert W \vert^5)  \nonumber \\
& & + {\pi^6 \over 1080} (Y^4 \vert W \vert^3 + \vert Y \vert^3 W^4),
\end{eqnarray}
\end{fleqn}
where $Y=N(y-x)/L$ and $W=N(w-u)/L$. To obtain Eq. (\ref{rho2_exp}) of the main text we have defined

\begin{equation}
  T_n=\sum_{m=0}^n A_{m,n} |Y|^m|W|^{n-m}.
\end{equation}

%---Appendix III -------
\section{Distribution of $N_0$ from spectral determinants}
\label{app_Bazhanov}
We briefly explain here how we have calculated the distribution of $N_0$ shown on Fig. \ref{fig_pno} and how the Gaussian behavior of the tail of the distribution $P(N_0)$ can be retrieved with the help of spectral determinants of Ref. \cite{Bazhanovetal}. Using the formulation of Ref. \cite{Gritsevetal1} (for a different problem of interferences between two interacting bosonic gases but mathematically similar to the problem considered in this article), the statistical properties of $N_0$ are related to the spectrum $\{\varepsilon_n\}$ of the radial sextic oscillator 
\begin{eqnarray}
- {d^2 \psi(r) \over dr^2}  + \Bigl(r^{6} + {\ell (\ell+1) \over r^2} \Bigr) \psi(r) \, = \varepsilon_n\, \psi(r),
\end{eqnarray}
with angular momentum $\ell = -{1 \over 2}$ and $r\in[0,+\infty[$. The distribution of $N_0$ is given by the following integral \cite{Gritsevetal1}
\begin{equation}
  P(\alpha)=2\int_0^\infty \prod_{n=1}^{\infty}\left(1-\kappa\frac{x^2}{\varepsilon_n}\right) J_0(2x\sqrt{\alpha})\,x\,dx,
\end{equation}
with $\alpha=N_0/\langle N_0\rangle$, $\kappa=8\sqrt{2}\,\Gamma(3/4)^2/\pi^2$ and $J_0$ the Bessel function. This is the result shown on Fig. \ref{fig_pno}.

The moments of the distribution can be cast in the form 
\begin{align}
  &\langle N_0^n \rangle  \equiv (\rho_\infty \sqrt{2N})^n\, Z_{2n} \nonumber \\
  & = (n!)^2 (\rho_\infty \sqrt{2N}\kappa)^n\!\!\!\!\! \!\!\!\!\!\!\!\!\!\!  \sum_{i_1,i_2,...,i_n, \, {\rm all}\,\,{\rm different}}\!\!\!\!\!\!\!\!\!\!  \prod \varepsilon_{i_1}^{-1}  \varepsilon_{i_2}^{-1} \cdots \varepsilon_{i_n}^{-1}.
\end{align} 
As explained in Ref. \cite{Bazhanovetal}, for $j$ larger than $2$ basically, $\varepsilon_j$ increases as $j^{3/2}$ and thus $Z_{2n}$ behaves as $\sqrt{n!}$ for large $n$. This in turn implies that $P(N_0) \simeq \exp(-C N_0^2)$ for large $N_0$, where $C$ is a real positive constant. The behavior of $\langle N_p^n\rangle$ for small but non-zero $p= j {2\pi\hbar \over L}$ is obtained in the same way, except that now, the energy levels $\varepsilon_i$ are no longer the energy levels of the oscillator in (C2) with $l=-{1 \over 2}$ but with $l = 4 j - {1 \over 2}$. For $j$ much smaller than $n$, the behavior of $\langle N_p^n \rangle$ still has the same behavior as $\langle N_0^n \rangle$, so the tail of the distribution $P(N_p)$ is also Gaussian. However, for $n$ much smaller than $j$,  $\langle N_p^n \rangle$ behaves as $n! \,\langle N_p \rangle^n$, signaling the exponential behavior of $P(N_p)$ for $N_p \ll j \langle N_p \rangle$.


\begin{thebibliography}{99}
\bibitem{Gumbel} In the weak interaction regime \cite{LovasDoraDemlerZarand}, $x=(N_0-\langle N_0\rangle)/\Delta N_0$, where $\Delta N_0$ is the standard deviation of $N_0$, is distributed according to the Gumbel distribution $P(x)=c\exp[c\,x-\gamma -\exp(c\,x-\gamma)]$, with $c=\pi/\sqrt{6}$, $\gamma\simeq 0.5772$ is Euler's constant.
\bibitem{noteint} The multi-dimensional integral in Eq. (\ref{eq_N0n}) with $n=2$ has been evaluated with a Monte-Carlo algorithm with $10^8$ points. The statistical sampling over ten thousand configurations gives $\langle N_0^2\rangle/\langle N_0\rangle^2=1.328$ with a standard deviation equal to $0.015$.
\end{thebibliography}
\end{document}